IAC-25-A6,IP,41,x94961

# Introduction to the Sky Survey Schedule (SSS) framework

Kai-Tian Yuan[a], Hou-Yuan Lin[a]*

[a] *Purple Mountain Observatory, Chinese Academy of Sciences, Nanjing 210023, China*
\* Corresponding Author

**Abstract**

To fulfill the requirements of space object cataloging and enable automated intelligent responses to anomalous events, we designed a novel observation scheduling system named Sky Survey Schedule (SSS). This framework facilitates coordinated operations across multi-site observational networks comprising dozens of instruments, while simultaneously supporting asteroid monitoring and time-domain astronomy studies. The system implements two principal observation modes: fixed sky regions and target-centered tracking. The former is used for sidereal or static observation, while the latter provides dedicated follow-up capabilities for transient targets. The sky regions are divided into latitude bands, each of which is subdivided into sectors to ensure minimal overlap and comprehensive coverage. These sectors are mapped to high-level HEALPix sky grids, enabling rapid cross-referencing and correlation between instruments. At the core of SSS lies an adaptive weighting architecture that integrates multiple parameters. Initial target priorities are determined from orbital catalogs containing both known and uncorrelated objects according to cataloging requirements. The system implements dynamic weight adjustments through feedback mechanisms: confirmed stable objects receive decaying weights, long-unobserved targets experience weight recovery, while anomalies (e.g., newly detected or lost objects) trigger priority escalation. These target-specific weights combine with observational factors - including phase angle constraints, lunar interference, Earth shadowing, and elevation limits - to generate space-time priority matrices. This quantitative framework systematically incorporates operator-defined priorities for specific regions/targets through configurable weight modifiers. Observation plans are dynamically optimized considering: 1) Instrument-specific constraints; 2) Site-level observational redundancy; 3) Weather-induced schedule disruptions; 4) Equipment availability. The system features an extensible API supporting customized observation requirements. For resource-rich sites (>100 deg² collective field coverage), an all-sky scan along the right ascension direction can be initiated.
**Keywords:** Space Situational Awareness, Optical observation, HEALPix, Sensor Task Scheduling

## 1. Introduction

With the rapid development of space activities, the near-Earth orbital environment has become increasingly complex. According to the statistics released by the European Space Agency (ESA) and the United States Space Surveillance Network (SSN), the number of trackable space objects has exceeded 40,000, including operational spacecraft, defunct satellites, and large rocket debris [1]. In addition, hundreds of thousands of smaller debris pieces are believed to exist. This exponential growth in space objects has posed severe challenges. On the one hand, the risk of collision with orbital debris has risen geometrically. On the other hand, objects undergoing frequent orbital maneuvers exhibit highly complex dynamics, making their position prediction subject to significant uncertainty [2]. Without real-time and efficient observation capabilities, the accuracy of orbit determination and the timeliness of collision warning would be severely compromised, threatening the safe operation of space assets.

Ground-based optical observation networks have emerged as an essential technical means for space situational awareness, owing to their unique advantages. Compared with radar and other active detection approaches, optical telescopes feature relatively low construction costs, flexible deployment, and wide fields of view (FOV). These advantages make them highly suitable for a variety of applications, including space debris monitoring, near-Earth asteroid warning, and time-domain astronomical survey [3]. A coordinated network composed of multiple telescopes not only enhances sky coverage and data acquisition efficiency, but also improves orbit determination accuracy through multi-station observations, while enabling rapid response to unexpected events. Nevertheless, conventional static scheduling modes suffer from three major limitations: (i) the lack of global optimization in observation task allocation, leading to low utilization of limited resources; (ii) delayed feedback mechanisms, which hinder timely handling of anomalous targets; and (iii) rigid scheduling strategies, which are unable to adapt to dynamic changes in instrument status or observation conditions [4]. These shortcomings severely constrain the overall effectiveness of space object monitoring systems.

The choice of mission planning and optimization methods depends largely on the objective function of the monitoring mission. In practice, the problem is particularly challenging because space debris moves at high velocities, exists in massive quantities, and is widely distributed across orbital space. Therefore, there is an urgent need to use limited network resources to generate optimized observation plans in near real time to meet mission requirements. To address these challenges, this paper presents Sky Survey Scheduler (SSS), an intelligent observation scheduling system designed for large-scale monitoring networks under multi-task and multi-instrument conditions. The system integrates priority evaluation of observation targets, dynamic resource allocation, sequence optimization, and real-time feedback, enabling the intelligent generation and adaptive adjustment of observation plans. It supports multiple observation modes, including wide-area debris surveys and precision tracking, while remaining compatible with additional scientific applications such as asteroid warning and time-domain astronomy. Through these features, this framework significantly enhances data collection efficiency, improves automation in schedule generation, and strengthens the overall capability of space debris monitoring and cataloging.

## 2. Functional model
### 2.1 Sky regions

This study adopts the Hierarchical Equal Area and isoLatitude Pixelization (HEALPix) scheme for the efficient management of a large number of space targets. This method ensures uniform pixel areas even in high-latitude regions and provides a hierarchical recursive division mechanism [5]. The resolution can be dynamically adjusted based on the number of telescopes, field of view size, and specific mission requirements, enabling seamless switching from large-scale to fine-scale regions. Assuming that the division level is $N_{\text{side}}$, the total number of pixels $N_{\text{pix}}$ is given by:

$$N_{\text{side}} = 2^k \quad (1)$$
$$N_{\text{pix}} = 12 \cdot N_{\text{side}}^2 \quad (2)$$

where k represents the number of layers. Figure 1 shows the pixels available in HEALPix when k = 4.

In this work, the sky regions are divided into latitude bands, each of which is subdivided into sectors to ensure minimal overlap and comprehensive coverage. The division is performed in the International Celestial Reference System (ICRS). The ICRS-based division avoids the reduction of the instantaneous visible regions caused by time drift, improving observation efficiency and task allocation stability. Figure 2 shows a schematic diagram of the sky regions based on a 7° × 7° FOV. where each HEALPix cell can be mapped to a specific sky region. The association with HEALPix pixels and their rapid retrieval enables quick matching across different observational sites and equipment (which may employ different sky division schemes). Based on the hierarchical structure of the HEALPix sky grids, target pixels can be quickly retrieved within a given region, greatly reducing the computational burden of large-scale spatial queries. This property is particularly advantageous for observation scheduling, as it enables rapid location of visible targets and real-time adaptation to dynamic observational constraints.

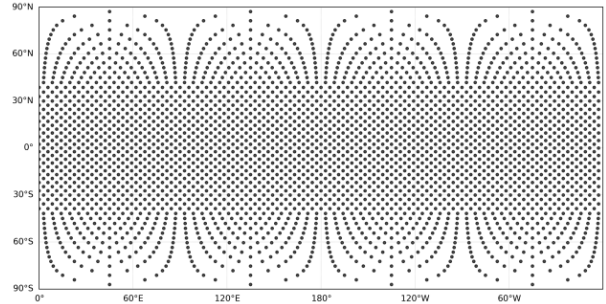

Fig. 1. The spherical HEALPix projection onto the plane when k = 4.

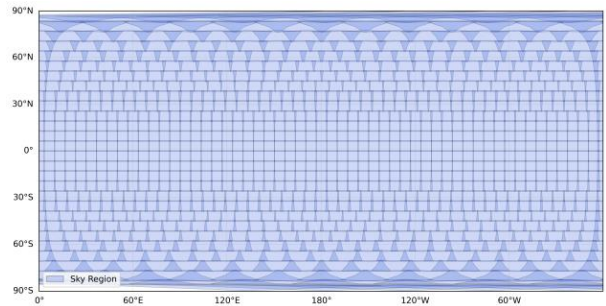

Fig. 2. Sky regions division based on the ICRS with the 7°×7° field of view.

For optical observation tasks, the visibility of a sky region is affected by multiple factors, including solar phase angle, moonlight, and elevation angle. To enable quantitative scheduling, we construct corresponding penalty functions for each factor. The solar phase angle $α$ is defined as the angle between the observation region and the Sun, which affects the target's illumination angle and thus its visibility. Meanwhile, during the twilight period, it also influences the sky brightness in the vicinity of the Sun. Hence, when the phase angle exceeds 90°, the weight is set to 1, while for smaller angles it decreases accordingly, which is:

$$\psi_{\text{SolarPhase}} = \begin{cases} 1 & , α \geq 90° \\ 1 - \cos α & , α < 90° \end{cases} \quad (3)$$

Additionally, based on the lunar calendar, we compute the avoidance angle $\varphi$ according to different moon phase and assign weights depending on the angular separation $\delta$ between the target region and the moon. Specifically, regions within the avoidance zone are fully suppressed with a weight of 0, while those within the attenuation zone are reduced to 0.3. That is,

$$\psi_{moonlight} = \begin{cases} 1 & , \delta >= 2\varphi \\ 0.3 & , \varphi \leq \delta < 2\varphi \\ 0 & , \delta < \varphi \end{cases} \quad (4)$$

However, these parameter values are preliminary and will be updated upon completion of a detailed follow-up study on lunar influence under various lunar phases and lunar angular distances [6]. For regions at low elevation, the effective weight is reduced to 60%. The elevation penalty function can be expressed as:

$$\psi_{lowalt} = \begin{cases} 1 & , \theta \geq \theta_{low} \\ 0.6 & , \theta_{min} \leq \theta < \theta_{low} \\ 0 & , \theta < \theta_{min} \end{cases} \quad (5)$$

where, $\theta_{min}$ represents the minimum elevation angle limit. Finally, the comprehensive benefit of a sky region is given as the product of these factors:

$$\psi = \psi_{SolarPhase} \cdot \psi_{moonlight} \cdot \psi_{lowalt} \quad (6)$$

By converting environmental constraints into quantitative penalty functions, different observation conditions can be directly compared within a unified framework. In addition, the weight factor ensures that severely degraded conditions are automatically suppressed, while favourable regions are prioritized. As a result, the scheduler can efficiently allocate limited observation resources to maximize scientific return while avoiding low-quality or infeasible observations.

*2.2 Target set filtering*

When facing tens of thousands of potential space targets, it is essential to refine the target set prior to scheduling in order to ensure both computational efficiency and the effectiveness of observation tasks. In this work, only those targets $i$ that satisfy a set of visibility constraints are retained during the optical prediction stage. Specifically, a minimum elevation criterion $C_{elev}$ is imposed to ensure that satellites are observable above a prescribed altitude threshold $\theta_{min}$. In parallel, the ground station must be between the evening and morning nautical twilight periods during the observation period $t$, requiring the solar altitude angle $h(t)$ must be lower than $h_{min}$, while the satellite must remain in sunlight rather than obscured within the Earth's shadow. To further ensure the feasibility of observations, the visible tracklet length $L_i$ in the telescope's field of view must exceed the specified minimum value $L_{min}$, and targets should be set to those whose expected apparent magnitude $m_i$ is within the instrument's detection limit $m_{lim}$. Accordingly, the candidate target set $T_{valid}$ can be formally expressed using set operations as the intersection of all constraints, which is:

$$\begin{cases} C_{elev} = \{(t,i) \mid \theta_i(t) > \theta_{min}\} \\ C_{night} = \{t \mid h(t) < h_{min}\} \\ C_{eclipse} = \{(t,i) \mid \varphi_i(t) < 90° \lor \sin\varphi_i(t) < R/r\} \\ C_{arc} = \{i \mid L_i > L_{min}\} \\ C_{mag} = \{i \mid m_i \leq m_{lim}\} \\ T_{valid} = C_{elev} \cap C_{night} \cap C_{eclipse} \cap C_{arc} \cap C_{mag} \end{cases} \quad (7)$$

where $\varphi_i(t)$ is the geocentric angle between the satellite and the Sun, $R$ is the Earth's radius, and $r$ is the geocentric distance of the target $i$.

Based on the existing orbital database, which includes cataloged objects and Unrecognized Targets (URTs), each target is initially assigned a base weight determined by cataloging and monitoring requirements. Targets with small or medium radar cross-section (RCS) values, which may be difficult to detect, are penalized by halving their effective weight. For URTs with orbital solutions older than a few days, the weight is further reduced due to increasing orbital uncertainty.

$$w_i(t) = \alpha_i \cdot w_{0,i} \cdot \gamma_i \cdot \rho_i(t) \quad (8)$$

where $w_{0,i}$ is the base weight, $\gamma_i \in (0,1]$ is a reliability factor reflecting RCS size or orbit staleness. $\rho_i(t)$ is the time recovery term parameter, and $\alpha_i \geq 1$ is an operator-defined priority coefficient from the task database.

The target weights are further adjusted through a dynamic updating mechanism driven by both observation feedback and orbit reliability. Target weights evolve over time through an adaptive feedback process: (i) weights of successfully observed and identified targets are reset to zero, (ii) weights of unobserved targets gradually recover, and (iii) newly detected, lost, or anomalous targets are assigned higher priority weights. Furthermore, operator-defined priority coefficient can be imported from the task database, allowing critical targets to recover weight more rapidly and thus shortening their revisit period. This formulation benefits observation scheduling by providing a flexible, feedback-driven mechanism that balances catalog maintenance, uncertainty reduction, and mission priorities. By adaptively updating weights, the system ensures that limited observation resources are allocated to the most valuable target, thereby improving both efficiency and long-term coverage.

*2.3 Search strategies*

Space objects frequently undergo orbit maneuvers or anomalous variations, which, if not promptly monitored, may lead to target loss or catalog corruption. In routine space object monitoring, observation tasks are typically guided by propagating historical orbital elements to predict target positions, which then direct optical or radar sensors for measurement. However, with the increasing frequency of orbit maneuver events, the orbit updating process encounters two principal anomalies. The first is target loss due to large maneuvers: significant deviations between the actual and predicted positions may cause the object to drift outside the observation FOV. The second is mismatched data association: newly acquired measurement arcs may not be reliably linked to existing cataloged orbits. To address these two types of anomalies, the SSS framework implements a directed search strategy and a feasible-region construction method based on dynamical constraints.

For the first case of large-scale deviations, a local search grid is generated around the predicted location, to enhance coverage for maneuvering targets. The observation scheduler then evaluates the sky regions corresponding to these points, selects the region with the highest comprehensive weight, and executes stationary observations to increase exposure time and enhance the detection probability of faint objects. For the second case involving unmatched observations, the system extracts associated arcs of unidentified targets and constructs feasible orbital regions under the two-body energy integral constraint for Earth-centered motion [7]. Based on the attributes of short-arc observations, the target can be constrained within a specific region.

By embedding both search-grid strategies for large deviations and feasible-region construction for unmatched data into the survey scheduler, the system can rapidly respond to abnormal events while maintaining global survey efficiency. This property benefits observation scheduler by enhancing the probability of capturing abnormal targets.

**3. SSS framework**

The core objective of the SSS framework is to construct a fully automated observation scheduler and execution framework for coordinated multi-telescope operations. The framework is designed to operate efficiently and reliably under complex observational conditions and multi-task competition environments, while maintaining scalability to accommodate large-scale survey missions. Driven by automated scheduler, the framework integrates multiple dimensions including target weights, visibility predictions, resource availability, and observation window matching, and employs optimization algorithms to automatically generate all-sky observation missions. Building on this foundation, a cross-site, multi-instrument coordination mechanism is implemented to achieve resource sharing and task synchronization. Additionally, conflict detection and path optimization effectively minimize scheduling conflicts and operational overhead. To enhance autonomy and adaptability, the SSS framework integrates a dynamic feedback loop. By continuously monitoring observation results and instrument status, the framework adaptively updates target weights and adjusts subsequent scheduling strategies, thereby achieving closed-loop control.

*3.1 Framework composition*

The SSS framework adopts a modular and extensible architecture, with each functional component remaining independent. This design allows for flexible replacement of algorithms and expansion of functionality to accommodate different survey requirements. Compared with conventional schedulers, the framework demonstrates superior efficiency in multi-instrument coordination, improved scheduling stability, and higher task completion rates. At the same time, it maintains a balance between automation and operator intervention, making it suitable for surveys of varying scale and strategy. The SSS framework is designed for automated scheduling and dynamic optimization of large-scale sky surveys. As shown in Figure 3, it consists of three main subsystems: the Target Ephemeris Calculation subsystem, the Target Weight Adjustment subsystem and the Mission Scheduler subsystem. To ensure data consistency and traceability, the SSS framework uses database-driven unified management, avoiding reliance on scattered files. The database layer includes Target Orbit Elements database, Sky Region database, Trajectory database, Station database, Device database, Target Weight database, Key Task database, and Observation Schedule database.

The SSS framework implements sky regions based on the HEALPix algorithm with configurable resolution and overlap. Each region is assigned center coordinates, boundaries, and cell indexes, as well as mapped to the instrument field of view for task assignment. The Target Ephemeris Calculation subsystem calculates the target position based on the orbit elements of the given time series, associates it with the sky region. It then filters valid observation windows under constraints such as elevation angle, sky background, earth shadow, arc length and magnitude. The target weights are initialized and synchronized with the database, while various coordinates and observation attributes are stored for use by the other modules. Target weights are determined jointly by base weights and the adaptive adjustment mechanism. The Target Weight Adjustment subsystem defines a recovery rate for weights, enabling unobserved targets to gradually recover their importance. Observation results are compared with scheduled tasks to dynamically reset observed targets, elevate the priority of anomalous or newly detected targets, and insert dedicated search tasks when necessary. For targets in Attention mode that have not been observed for longer than the set duration, the framework will automatically add search tasks to the Key Task database for enhancing coverage of maneuvering targets. Feedback results, target weights, device status, and other information are also synchronized with the database to ensure consistency across modules.

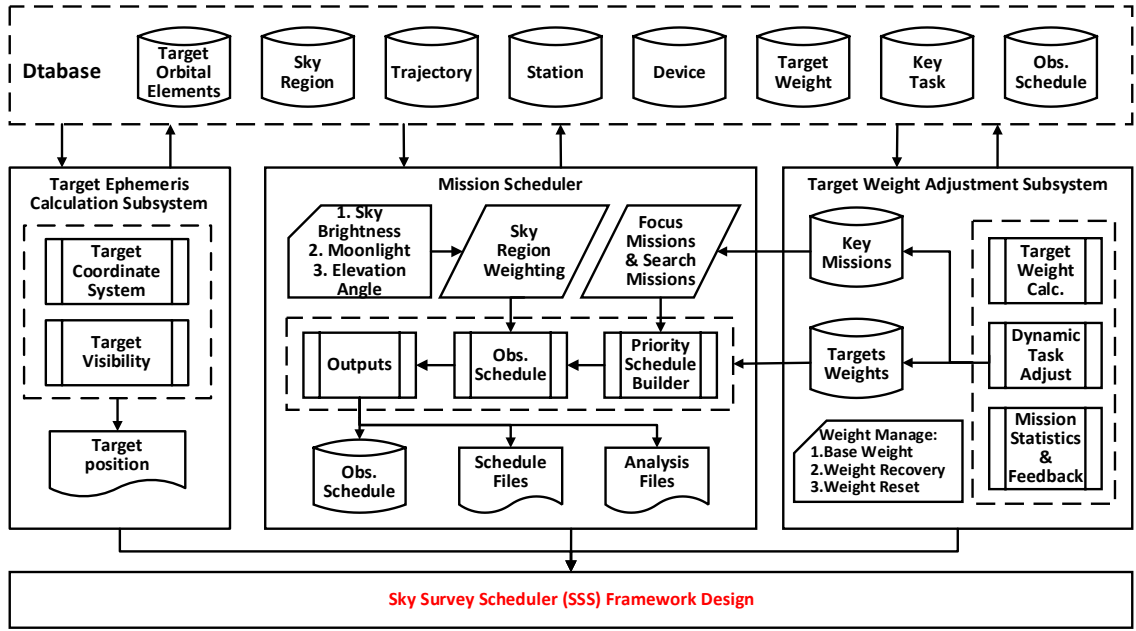

Fig. 3. Design of the Sky Survey Scheduler (SSS) framework.

*3.2 Task types*

In the process of space targets monitoring, timely and prioritized tracking of high-value or potentially dangerous objects is critical to reducing collision probabilities and preserving catalog integrity. The SSS framework adapts to a wide range of scientific missions by defining multiple observation modes and mission categories, thereby achieving comprehensive coverage from routine sky surveys to rapid response for key sky regions. Within the SSS framework, tasks are classified into several types, each corresponding to specific operational modes and observation strategies. These task types are summarized in Table 1. The observation strategy revolves around two principal modes: Attention mode and Focus mode. Additionally, the framework integrates Search mode and Virtual mode as supplementary modes, forming a comprehensive task system. Users can submit priority tasks to the Key Task database using target IDs or coordinate coordinates. The orientation task submissions support coordinate systems such as ICRS, International Terrestrial Reference System (ITRS), and the horizontal coordinate system. Observation modes encompass Sidereal mode, Static mode, and Guide mode for adaptation to different types of tasks.

Table 1. Common task types in the Key Task database

| Task type | Task mode | Observation mode |
|---|---|---|
| Target | Attention | Sidereal/ Static/ Guide |
|  | Focus | Sidereal/ Static/ Guide |
|  | Search | Sidereal/ Static |
| Orientation | Attention | Sidereal/ Static/ Guide |
|  | Focus | Sidereal/ Static/ Guide |
| Virtual | Attention | Sidereal/ Static |

In practical applications, the attention mode achieves adaptive re-visits of priority targets within defined time intervals through dynamically updated weights and pre-defined sky regions. The focus mode ensures that high-priority targets are placed at the center of the FOV during critical time windows, thereby supporting continuous tracking tasks for specific targets. However, in actual tasks, to meet cataloging requirements, specific targets need to be revisited and tracked at a fixed time frequency. Therefore, the Attention Guide mode is added, which optimizes task allocation based on the required revisit cycle using Gaussian-weighted time and places the target at the center of the field of view. This mode is used for rapid search and localization of unknown or anomalous targets and is often used in conjunction with the Attention mode. For targets in Attention mode that have not been observed for more than the set duration, the framework automatically adds the search task to the Key Task database to enhance coverage of maneuvering targets. Then, the Mission Scheduler subsystem reads the Key Task database, calculates the weight for the relevant sky region, and adds it to the candidate sequence along with other sky regions. Based on dynamic weights, the scheduler selects the optimal sky region for observation. The process for adding search tasks is shown in Figure 4.

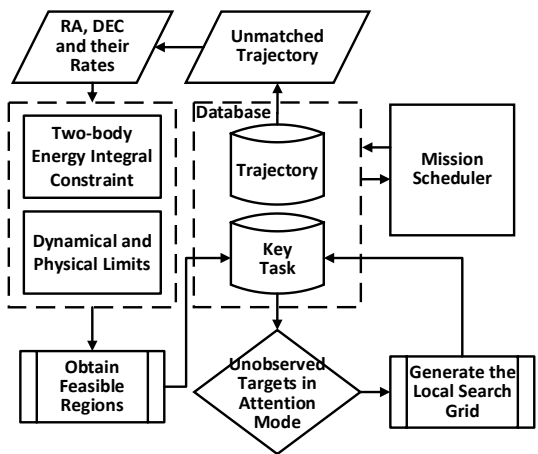

Fig. 4. Flowchart of adding search missions to the Key Task database.

This framework also introduces the Virtual mode, extending observational capabilities beyond cataloged celestial objects to arbitrary points of interest. Virtual points can be defined as fixed locations or moving targets and support multiple coordinate systems. Each virtual point is assigned an initial weight determined by the operator, after which its weight is dynamically updated along with the average score of all targets within the corresponding sky region. This design ensures that the weight of a virtual point is fully compatible with the existing weighting structure, allowing both target and virtual weights to be integrated into a unified evaluation of sky regions. The scheduler selects the optimal observation region based on these dynamic weights. By enabling the inclusion of arbitrary sky positions as observation candidates, the Virtual mode proves particularly valuable in applications such as fragmented debris searches or time-domain astronomical observations.

These task types collectively establish a unified operational framework that balances sky surveys with precise tracking, enabling the system to seamlessly adapt to different scientific and surveillance objectives. This hierarchical structure provides significant benefits for observation scheduler by enabling flexible task priority settings, efficient resource allocation, and rapid response to evolving observation requirements.

### 3.3 Mission scheduler

The Mission Scheduler subsystem is the most significant component of the entire SSS framework, responsible for orchestrating observation plans. The processing flow is shown in Figure 5. The scheduling process begins by loading instrument metadata and determining its operational states, while synchronizing with the Device database. Non-operational instruments are automatically assigned synchronous belt sky survey tasks to ensure coverage and facilitate calibration, while active instruments proceed to the standard scheduling sequence.

First, based on the other two subsystems, the target set and sky region are initialized. Based on observational constraints, including solar phase angle, moonlight, elevation angle, etc., fixed sky region weights are calculated. Combined with the determined dynamic target weights, this forms a two-dimensional weighting system, reflecting the priority of each sky region at each time. This weighting structure forms the basis for developing observation plans, and the priorities defined by operators for specific regions or targets are also quantitatively incorporated into the framework.

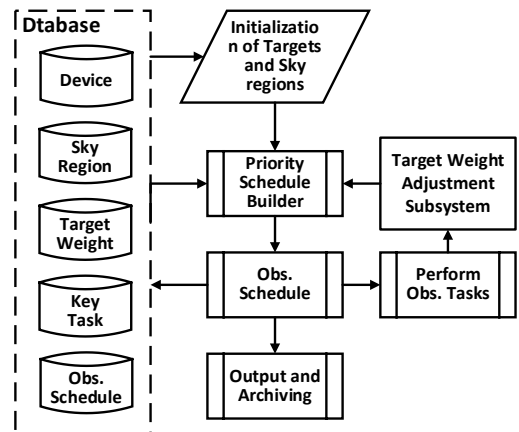

Fig. 5. Flowchart of the Mission Scheduler subsystem.

After completing the reading and processing of basic information, the scheduler executes priority missions in the Key Task database. Among them, the center-pointing task is scheduled first. The scheduler performs visibility analysis based on the target position or pointing information and then schedules it into the specified time window. Search mode tasks enter the candidate area sequence and are evaluated together with other regions based on dynamic weights. Attention mode tasks increase the base weight of the specified target and shorten its re-visit cycle. Then, the scheduler selects the appropriate sky region based on two-dimensional weights and completes the scheduling of the remaining regular tasks. The scheduler optimizes task allocation among multiple instruments within the single station to reduce path redundancy and avoid resource conflicts. Observation plans will also be dynamically adjusted in response to missing observation results caused by weather or equipment failures. The final plan is executed, and the results are fed back to the Target Weight Adjustment subsystem, after which the mission is adaptively updated in a closed-loop cycle. Finally, synchronize the generated task sequence with the Observation Schedule database and generate local files for subsequent analysis and archiving. The integration of a two-dimensional dynamic weight structure with modular priority handling ensures that scheduling decisions balance survey completeness, anomaly response, and resource efficiency. By combining static constraints with adaptive feedback, the framework maintains high responsiveness to unexpected event while preserving the overall stability of long-term survey coverage.

### 4. Operation and Discussion

Here we demonstrate the operational execution of the observation plan using a network of 12 telescopes located at the Yao'an Observatory in Yunnan province, China (long. E 101°10'54.98", lat. N 25°31'41.02", alt. 2014.4 m). The experimental setup is summarized in Table 2, which specifies the key experimental parameters, including the telescope FOV, temporal resolution, and visibility constraints. In addition to general survey missions, three priority targets were introduced to evaluate the system's capability in handling high-priority monitoring modes. Specifically, the Attention mode was assigned to BeiDou-10 (C10) for continuous surveillance across multiple intervals, while the Focus mode was applied to BeiDou-56 (C56) for stringent tracking during critical time windows. Concurrently, BeiDou-7 (C7) was observed using the Attention Guide mode to ensure tracking with a fixed revisit cycle. These configurations allow for a comparative assessment of the system's ability to integrate both routine surveys and specialized monitoring tasks within the same scheduling framework.

Table 2. Summary of experimental settings and processing strategies.

| Parameter | Value |
|---|---|
| Number of telescopes | 12 |
| Time window (s) | 120 |
| Minimum elevation angle | 12° |
| Telescope FOV (deg²) | 7 × 7 |
| Maximum longitude overlap | 1° |
| Key targets | C7, C10, C56 |

For sky division, each telescope was assigned a fixed region scheme with a 7° × 7° FOV. This division scheme allows a maximum longitudinal overlap of 1° to minimize redundancy while ensuring continuous coverage. For each cell, we computed the center coordinates, boundaries, and included HEALPix cells. The generated set of 415 instantaneous visible regions in ITRS is shown in Figure 6. These regions evolve over time due to Earth rotation and orbital geometry, while the background sky remains fixed. In the figure, the green dashed line indicates the 12° minimum elevation angle, and the black triangle marks the station. The target catalog consisted of 4,164 medium- and high-Earth orbit (MEO/HEO) objects, whose detailed composition is summarized in Table 3. Based on the visibility analysis shown in Figure 7, 3,040 objects were observable during the observation period, where green dots in the figure denote the visible subset. For each time step, the system calculated the position of every target, mapped them to their corresponding HEALPix cells, and associated them with specific sky regions, thereby ensuring efficient spatial indexing and task assignment.

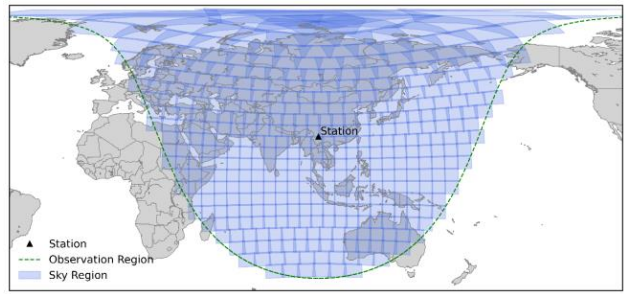

Fig. 6. Instantaneous sky coverage in ITRS.

Table 3. Target catalog attributes

|  | L | M | S | # |
|---|---|---|---|---|
| PAYLOAD | 1685 | 124 | 7 | 77 |
| R/B | 976 | 115 | 3 | 83 |
| DEBRIS | 211 | 472 | 201 | 186 |
| TBA | 3 | 0 | 0 | 21 |

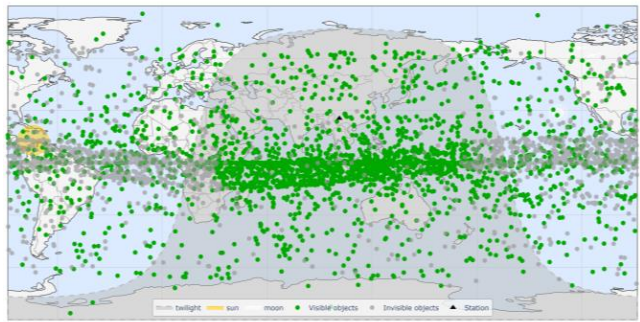

Fig. 7. Visibility analysis of the selected target set.

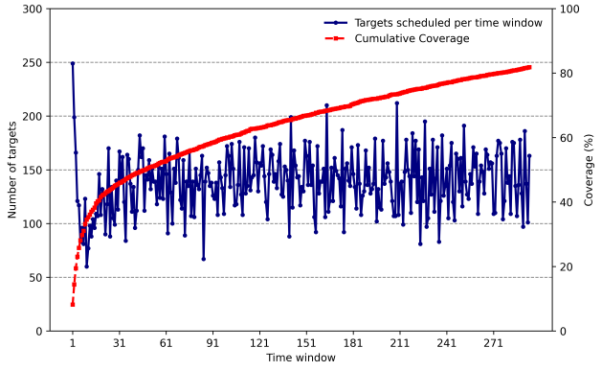

Fig. 8. Number of satellites scheduled per time window and cumulative coverage.

After applying visibility filtering, weighting initialization, and task assignment, the scheduler produced a final observation plan consisting of 2,488 scheduled targets and 41,319 observation arcs. Figure 8 illustrates the temporal distribution of scheduled target observations and their corresponding cumulative coverage rates. Each blue point represents the number of targets assigned to a specific time window across all devices, while the red curve depicts the cumulative coverage percentage relative to the total target count of all visible targets. This approach achieves broader search coverage and higher search efficiency while ensuring coverage and repeatability in key celestial regions. Figure 9 displays the completed observation schedule, where two representative time windows are selected to illustrate the spatial coverage of the sky regions. In the figure, red markers denote missions scheduled at the current time step, blue markers indicate tasks from the previous step, squares correspond to fixed sky-region survey tasks, and circles represent central pointing tasks. The three priority targets are highlighted with orange circles.

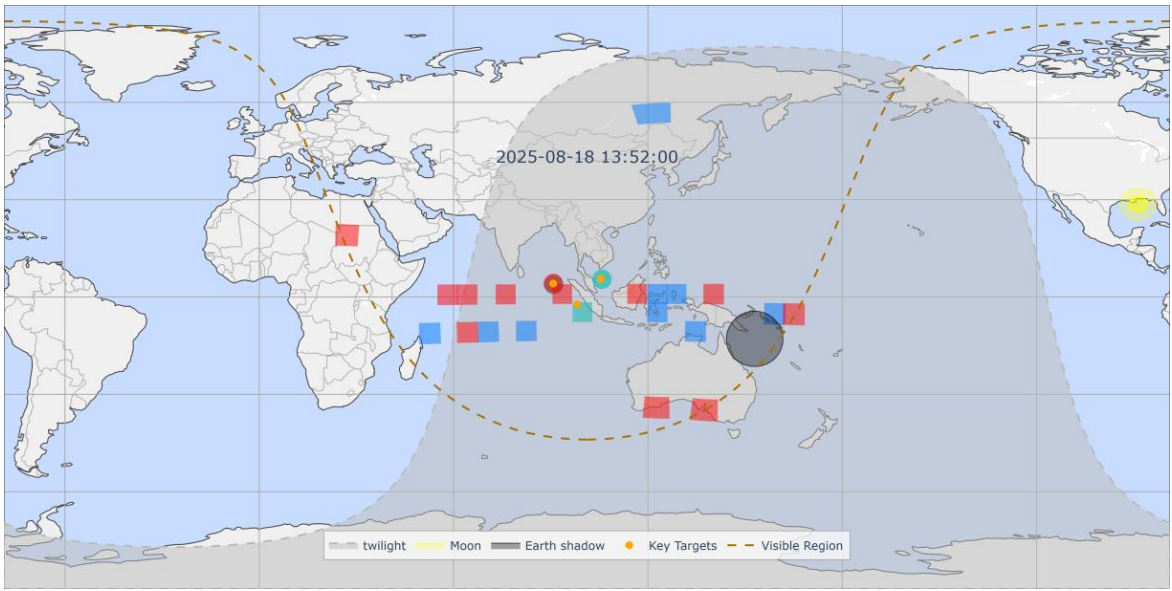

Fig. 9. Final observation schedule in the ITRS frame at GEO altitude, showing sky coverage across two time windows.

A closer examination of the observation modes reveals that central pointing tasks, including Focus and Attention Guide, successfully positioned priority targets at the center of the FOV. For example, under the Focus mode, C56 was continuously tracked throughout its designated time window, ensuring uninterrupted high-intensity monitoring. Under the Attention Guide mode, C7 was observed with a fixed revisit cycle of approximately 1.5 hours, resulting in six total observations with an average interval of 76.4 minutes. In contrast, fixed-region survey tasks did not necessarily center the targets within the FOV but instead followed the pre-defined sky regions. As a result, C10, whose priority weight was adaptively elevated, was observed 15 times within the observation period, with an average interval of 29.3 minutes, despite occasionally being captured near the FOV edge. These results demonstrate that the system is capable of balancing large-scale survey requirements with dynamic priority adjustments. The simultaneous execution of fixed survey tasks and specialized monitoring tasks highlights the adaptability of the scheduling framework, while the detailed observation statistics validate its capacity to ensure both coverage completeness and priority target responsiveness.

## 5. Conclusion and Future Work

In the proposed framework, three subprograms operate synchronously to enable real-time adjustment and reset of target weights. A key innovation is the integration of a HEALPix-optimized spatial indexing scheme, which facilitates efficient sky partitioning and thereby improves multi-instrument coordination. Furthermore, a closed-loop weight adaptation mechanism is employed to continuously update priorities based on observational feedback, which enhances responsiveness to critical targets. In addition, a probabilistic resource allocation strategy is introduced to balance automated scheduling with human oversight, ensuring both robustness and adaptability in complex survey environments. By continuously monitoring real-time conditions, fault-tolerant scheduling is achieved, enabling the system to maintain stable performance under uncertainties such as instrument downtime or observational failures. These features collectively significantly enhance observational efficiency while preserving the flexibility required to support specialized objectives.

While the SSS framework can effectively allocate observation tasks, several avenues for future research remain. First, enriching the analysis of observational outcomes will enable a more detailed understanding of system performance and provide refined feedback for weight adjustment mechanisms. Additionally, expanding the framework to support multi-site collaborative observation and data fusion will significantly improve coverage and resilience. Third, introducing quantitative evaluation of ultimate observation goals will provide a stronger scientific basis for prioritization and resource allocation. Finally, integrating reinforcement learning methods holds promise for optimizing scheduling strategies and dynamic weight adaptation in more complex and uncertain operational environments.